# Challenges to Bohr's Wave-Particle Complementarity Principle

**Mario Rabinowitz**

**Abstract** Contrary to Bohr's complementarity principle, in 1995 Rabinowitz proposed that by using entangled particles from the source it would be possible to determine which slit a particle goes through while still preserving the interference pattern in the Young's two slit experiment. In 2000, Kim et al used spontaneous parametric down conversion to prepare entangled photons as their source, and almost achieved this. In 2012, Menzel et al. experimentally succeeded in doing this. When the source emits entangled particle pairs, the traversed slit is inferred from measurement of the entangled particle's location by using triangulation. The violation of complementarity breaches the prevailing probabilistic interpretation of quantum mechanics, and benefits Bohm's pilot-wave theory.

## 1 Introduction

Thomas Young's 1802 two-slit interference experiment [1], in a myriad of variations, remains the archetype to this day for illustration and consideration of the issues related to wave-particle duality. It took only four years after Einstein explained the photoelectric effect [2] by quantization of light, for the first single-photon interference experiment to be performed. Taylor did it in 1909 with a flame light source, a diffraction grating and a photographic plate [3]. Taylor achieved the penultimate (considering delayed choice or quantum erasure to be the ultimate) double-slit experiment, in which particles are directed one at a time at a grating. These single particles are each diffracted in passing through a slit. Each particle produces one displaced spot on a screen opposite the slits; and with the collection of a large number of particles an interference pattern emerges.

Mario Rabinowitz
Armor Research; 715 Lakemead Way; Redwood City, CA 94062-3922 USA
e-mail Mario715@gmail.com

## 2 Background

In 1927 Niels Bohr [4] formulated the complementarity principle (CP) epistemologically, rather than mathematically, as a generalization of the newly devised mathematical Heisenberg uncertainty principle. Bohr taught that it is impossible to have a sharp distinction between the behavior of a system and the instruments used to measure it. Bohr's general statement of the CP is that information obtained under different experimental conditions cannot be comprehended within a single picture, but must be regarded as complementary in the sense that only the totality of the phenomenena exhausts the possible information about the objects.

As applied to wave-particle duality exemplified by Young's double-slit experiment, the complementarity principle of Niels Bohr [4] concludes that the interference pattern will always be destroyed when a determination is made of which slit is traversed by the quantum particle -- even when the uncertainty principle has not been violated. This was generally accepted as no experiment had ever contradicted the complementarity principle.

For electromagnetic radiation, the interference pattern can be understood both classically and quantum mechanically. Classically, the interference fringes arise from constructive and destructive interference at the observation screen of a spatially coherent electromagnetic wave going through the two slits. There are a number of quantum mechanical expositions such as the Born probability interpretation [5], the Heisenberg matrix mechanics [6], the Feynman particle quantum electrodynamics [7], and Bohm's [8] hidden variable approach to de Broglie's pilot-wave guiding particles in the direction of its interference after going through the two slits. Though differing considerably in their modeling of the experiment, they all give the same result for the interference pattern. However it appears that until 2012, only the pilot-wave model seemed to allow for determination of the traversed slit without erasure of the interference pattern.

Without a bias for one model over another, in 1995 Rabinowitz [9] predicted that it would be possible to ascertain the traversed slit, and still preserve the interference pattern. He proposed a novel double slit experiment in which the source emits entangled particle pairs. One of the particles (signal particle) *B* moves in the direction of the two-slit plate, and the partner (idler) particle *A* goes in the opposite direction towards a solitary luminescent screen *A*. [The term luminescent screen was used for any detector capable of single particle detection such as an array of photomultipliers.] He concluded that the traversed slit can then be determined from measurement of the entangled idler particle's position on the solitary luminescent screen *A*, the location of the source, and simple triangulation without violating the uncertainty principle. Experiments should be conducted to resolve whether this determination can be made before, at the same time, or after the slit is traversed by particle A depending on whether the ratio $D_A / D_B < 1, = 1, or > 1$. If the interference pattern is destroyed for certain values of this ratio, this may give insight into the entanglement process itself. As shown in Fig. 1, $D_A$ is the distance from the source to the solitary screen *A* and $D_B$ is the distance from the source to the two-slit plate which is in front of the interference screen *B*.

Thus the traversed slit is determined without in any way interacting with the two-slit plate, the interference luminescent screen *B*, or the interference particle *B*. All prior actual, and gedanken experiments of Einstein, Feynman, et al. that tried to determine the traversed slit and maintain the interference pattern, suffered from interaction with at least one of these elements.

## 3 Menzel et al. Experiment

In 2012 Menzel et al. [10] reported that they had successfully conducted a Young's two slit experiment in which they had located which slit a photon passed through, and contrary to Bohr's complementarity principle still obtained an interference pattern. Both references [9] and [10] utilized a source emitting entangled particle pairs. Both determined the traversed slit taken by the interference particle by observing the entangled particle. The interference particle was designated the "signal" particle and the observed particle was designated the "idler" particle by [10]. Whereas [9] called the observed particle, the "partner" particle.

As said on p. 779 of [9]: "we may determine … [what] subsets produce an interference pattern." Analysis of subset data in the Menzel et al paper [10] may be illuminating. Knowing the traversed slit permits obtaining single slit data without covering up the other slit. It is interesting to ascertain the kinds of interference patterns that emerge from their data for each slit separately, and for the sum of the two slit fringe data. Menzel et al make these statements regarding their data:

> Menzel et al [10] p. 9315 say "The evaluation of the experimental data yields a contrast ratio of about 1% cross correlation photons. Thus, from all the photons measured behind the slit in coincidence with the reference detector, at least 99% were passing through the related slit (upper-upper or lower-lower), and only 1% in maximum passed through the other slit (upper-lower or lower-upper) as a result of measurement errors. This experimental result confirms that photons measured in coincidence with the two selected positions of a reference detector are passing with at least 99% probability just through the one selected slit."
>
> Menzel et al [10] p. 9317 "In the near-field configuration the only atoms satisfying the coincidence condition are half-in-between the detectors and therefore sit on the node of the pump mode. Because the only atoms that could cause the appropriate clicks are not excited perfect correlations between signal and idler photon on the upper-upper and lower-lower slits emerge."

The Menzel et al subset data is sparse. It is nevertheless noteworthy. The main contribution to the interference is that resulting from the related slit traversed by the signal quantum particle. According to the pilot wave model, with both slits open, an interference pattern should emerge even though each slit is traversed separately. This results from interference from the "empty wave" through one slit with the "particle carrying wave" through the other.

### 3.1 Presence or Absence of Mechanical Slits

There is equivocation in two places (p. 9314 and p. 9316) about the use of mechanical slits in Menzel et al. [10]. On p. 9314, their abstract creates an ambiguity as to whether mechanical slits were actually used when they say, “According to quantum field theory the signal photon is then in a coherent superposition of two distinct wave vectors giving rise to interference fringes analogous to two mechanical slits.” [10] The text of their paper describes the presence of two mechanical slits,

A novel feature of the Menzel et al. experiment was the choice of the laser $TEM_{01}$ pump mode in the generation of their signal and idler photons. For them this is analogous to having two mechanical slits. Let us see how no-mechanical slit interference has been previously employed since a number of previous experiments did not use mechanical slits. Here are two such cases without mechanical slits that gave rise to interference fringes.

In 1993, Eichman et al. used Hg ions [11] as an analog of the two slits in Young's experiment. Light, both elastically and inelastically, scattered from the two ions produces an interference pattern. They used Doppler laser cooling to trap two $^{198}Hg^{+}$ ions at separations of 3.7, 4.3, and 5.4μ. These trapped ions scatter light of 0.194μ wavelength both elastically and inelastically to produce an interference pattern. They used polarization-sensitive detection of the scattered photons to switch on either the wave-like or the particle-like character of the scattered photon.

Quantum mechanics predicts that interference results from the scattered light when it is not possible at each collision to determine which ion scattered the photon. Quantum mechanics also predicts that no interference pattern will result when the adjacent atom is distinguished from the scattering atom because this allows the photon path to be determined. This is what Eichman et al. observed when they distinguished the atom’s internal level structure. When they determined the photon trajectory they obtained a particle-like behavior. They obtained the usual interference pattern when they didn't determine the photon path.

In 1985, P. Grangier et al. [12] conducted an experiment without mechanical slits, and evidence for an interference pattern that can be turned on or off. They observed modulation in the time-resolved atomic fluorescence light following the photo-dissociation of $Ca_2$ molecules. They said that "This modulation is due to an interference effect involving two atoms recoiling in opposite directions, while only one photon is emitted. ... This experiment is thus analogous to a single-photon Young's [double] slit experiment, in which the 'slits' (the atoms) are moving."

They ascertained the photon trajectory by observing the momentum of each atom after the photon is detected. This determines which atom (slit) was the emitter since it received the extra momentum $hn/c$ (where $n$ is the emission frequency) from the fluorescence photon. The difference in momentum of the two atoms needs to be observed with an uncertainty less than $hn/c$. But if the momentum uncertainty is less than $hn/c$, the uncertainty principle implies that the uncertainty in the separation of the two atoms must be greater than the emission wavelength. This destroys the interference effect. However, the initial dispersion of the relative position of the atoms is very small compared to the emitted wavelength, whereas the momentum difference dispersion is greater than $hn/c$ for these atoms with recoil velocity of ~ 500 m/sec, so that there is no way to know "which atom emitted the photon".

## 3.2 Origin of Entangled Quantum Particle Pairs From the Source

Entangled quantum particle pairs from the source have been crucial to determining the path taken from the source to one of the slits while preserving the interference pattern. For this reason, it is important to trace the origin of this concept. In a paper submitted 2/1/95 and published 6/10/95, Rabinowitz [9] proposed that measurement of the position of the partner (idler) particle determines which slit the signal particle traversed.

In 2000, Kim et al [13] employed spontaneous parametric down conversion to prepare entangled photons as their source, instead of using an atomic cascade decay process. The complete abstract of their published paper says, "We report a delayed 'choice' quantum eraser experiment of the type proposed by Scully and Drühl (where the 'choice' is made randomly by a photon at a beam splitter). The experimental results demonstrate the possibility of delayed determination of particlelike or wavelike behavior via quantum entanglement. The which-path or both-path information of a quantum can be marked or erased by its entangled twin even after the registration of the quantum." Interestingly the abstract of their 1999 ArXiv:quant-ph / 9903047v1 says "… The experimental results demonstrated the possibility of simultaneously observing both particle-like and wave-like behavior of a quantum via quantum entanglement…" If it had been "**simultaneously**," they would have been the first to invalidate complementarity. However, evidently it was later decided that they had not done this "**simultaneously**."

In 1982, Scully and Drühl [14] did not suggest the use of entangled particles in any aspect of their proposed experiment. In 2012, Menzel et al [10] reported an experiment in which they inferred the traversed slit by means of the entangled particle and still maintained the interference pattern. Although Menzel et al referenced Kim et al, they did not mention that the 2000 Kim et al reference used entangled particles. The 1995 reference [9] using entangled particles from the source was neither referenced by Menzel et al nor by Kim et al. Kim et al did not determine concurrent particlelike and wavelike properties as proposed by Rabinowitz [9] and observed 17 years later by Menzel et al.

Menzel et al [10] clearly state on p. 9314 that the entangled particles are crucial to ascertaining the traversed slit: "In the present paper we report the results of a double-slit experiment that brings out an additional layer of this principle. We employ the entanglement between the signal and the idler photon created in spontaneous parametric down-conversion (SPDC) (17) to obtain by a coincidence measurement of the two photons which-slit information about the signal photon without ever touching it. Moreover, we observe in a separate coincidence experiment interference fringes in the signal photon."

### 3.3 Critique of the Menzel et al. Experiment

Although Menzel et al. have done a difficult, impressive, and groundbreaking experiment, it is not beyond legitimate criticism from potential doubters. There is a curious statement on p. 9316 of the Menzel et al. paper [10]: “This analysis shows that the interference fringes in Fig. 4 are a consequence of the $TEM_{01}$ mode. Therefore, one might wonder if the mechanical double-slit is even necessary, especially because the one of the two slits is matched to the separation of the two intensity maxima of the mode. Moreover, our theoretical analysis does not contain the slits and we still obtain the fringes. Therefore, interference may be observable even without the double-slit but a substantial loss in contrast may occur.” This is both important and readily done.

It would have been an easy matter to simply remove the slit plate and see, as they suggested in their last sentence, that the interference pattern persists in spite of *welcher weg* (which path i.e. which lobe) knowledge. They say that their quantum field theoretical analysis does not contain the slits and they analytically still obtain the fringes. As mentioned in the early part of this section, there is a body of prior experimental work that used non- mechanical slits, and still obtained interference fringes. So it need not be in doubt that the observed interference may have more to do with the two-lobe structure of the $TEM_{01}$ mode than with the slits. What they have left in doubt for dubious readers is the possibility that the interference pattern survives a "which slit" determination, because as Menzel et al say, it has little or nothing to do with "which slit". A better test of wave-particle duality would have been a "which $TEM_{01}$lobe" experiment, that skeptics would expect to erase the interference pattern. Skeptics might assert that the "$TEM_{01}$ pump mode" introduces an additional superposition of wave vectors independent of the two slits. However, one may expect a similar result with a "which $TEM_{01}$lobe" determination, as they obtain theoretically.

There is a puzzling aspect in which two statements appear inconsistent with each other. Starting at the bottom of p. 9317 and ending at the top of p. 9318, Menzel et al [10] say: “However, we emphasize that even in our experiment the measurements of the which-slit information and interference still require mutually exclusive experimental arrangements. In order to observe the near-field coincidences shown in Fig. 3 the detector D2 has to be just behind one of the slits, whereas the interference fringes of Fig. 4 emerge only when we move D2 far away from the slits." Needing to "move D2 far away from the slits" seems to imply that the idler particle has to be observed before the signal particle to see interference. But they don't say this. The relative distances, and hence time sequences for measuring the idler photon relative to the signal photon appears not to be given, nor is it plain why they require "mutually exclusive experimental arrangements" of this kind. This is a clear distinction with the 1995 proposed experiment of [9] which discusses the consequences of measuring the idler photon before, at the same time, or after the signal photon; and does not require coincidence between the interference (signal) particle and the partner (idler) particle since this experiment can be conducted with particle pairs emitted one-at-a-time. Their own statement on p. 9314 appears inconsistent with the mutual exclusivity: "Moreover, the intensity of the pump beam is low enough to ensure that only one photon pair was created at a measuring time interval." This should obviate the need for coincidence counting that is the origin of their "mutually exclusive experimental arrangements." The relative positioning of the detectors may give insight into the entanglement process, rather than just relating to the complementarity principle.

Menzel et al [10] on p, 9318 go on to make the bold assertion: "In this sense our experiment can also be interpreted as another confirmation of the nonobjectifiability of quantum mechanics or, as stated by Torny Segerstedt, 'Reality is theory.'" Mirell [15] argues that locally real states of photons and particles are in exact agreement with quantum mechanics. Objectivity and consistency rather than subjectivity is the goal of science where one cannot talk about reality without observations or measurements. Although it is true that measurements or experiments are theory-laden, we normally know reality through experience rather than theory. It is not unreasonable to contemplate that not all prevailing theory has a bearing on reality, and that the combination of limited experiments and such theory may only serve to ascertain the relative consistency of our representations of nature.

After successfully conducting an experiment that preserves the interference pattern while determining the traversed slit by means of an idler particle, one more step would provide completeness and instill further confidence in the reality of the experimental results. Even though it will erase the interference pattern, an intrusive method of ascertaining the traversed slit should be done to validate the idler particle approach.

## 4 Uncertainty Principle

Since Menzel et al. [10] did not present an uncertainty principle analysis, it may be insightful to present one here. Let us ascertain whether or not the uncertainty principle is violated in determining the traversed slit by measurements on the idler particle $A$. Consider a quasi-stationary source that emits a particle-pair with an uncertainty $dx = \Delta x$ in the lateral x-direction parallel to the slit-plate, where $\Delta x$ may be considered to be the diameter of the source region in Figs. 1 and 2. The uncertainty principle gives the uncertainty in lateral momentum of the emitted particles as

$$dp_x \geq \hbar / 2\Delta x \tag{1}$$

Since the momenta of the entangled particles is equal and opposite, the lateral displacement of particle B at the slit plate is

$$\Delta s_x = \left( \frac{dp_x}{p_y} \right) D_B \ , \tag{2}$$

where py is the component of momentum perpendicular to the slit plate, and $D_B$ is the distance from the center of the source to screen B. To a good approximation

$$h / l = p \approx p_y \ , \tag{3}$$

where $l$ is the de Broglie wavelength of particle B.

Thus there is effectively a virtual beam from each spot on screen A to the slit plate. Consider a spot that is centered above the source and is small compared with the source width $\Delta x$. Fig. 2 illustrates this case. Other cases can be similarly analyzed. We want the beam spread at the slit plate from a given spot on screen A to be small compared with the slit separation,

$$s > \Delta x + 2\Delta s_x \ , \tag{4}$$

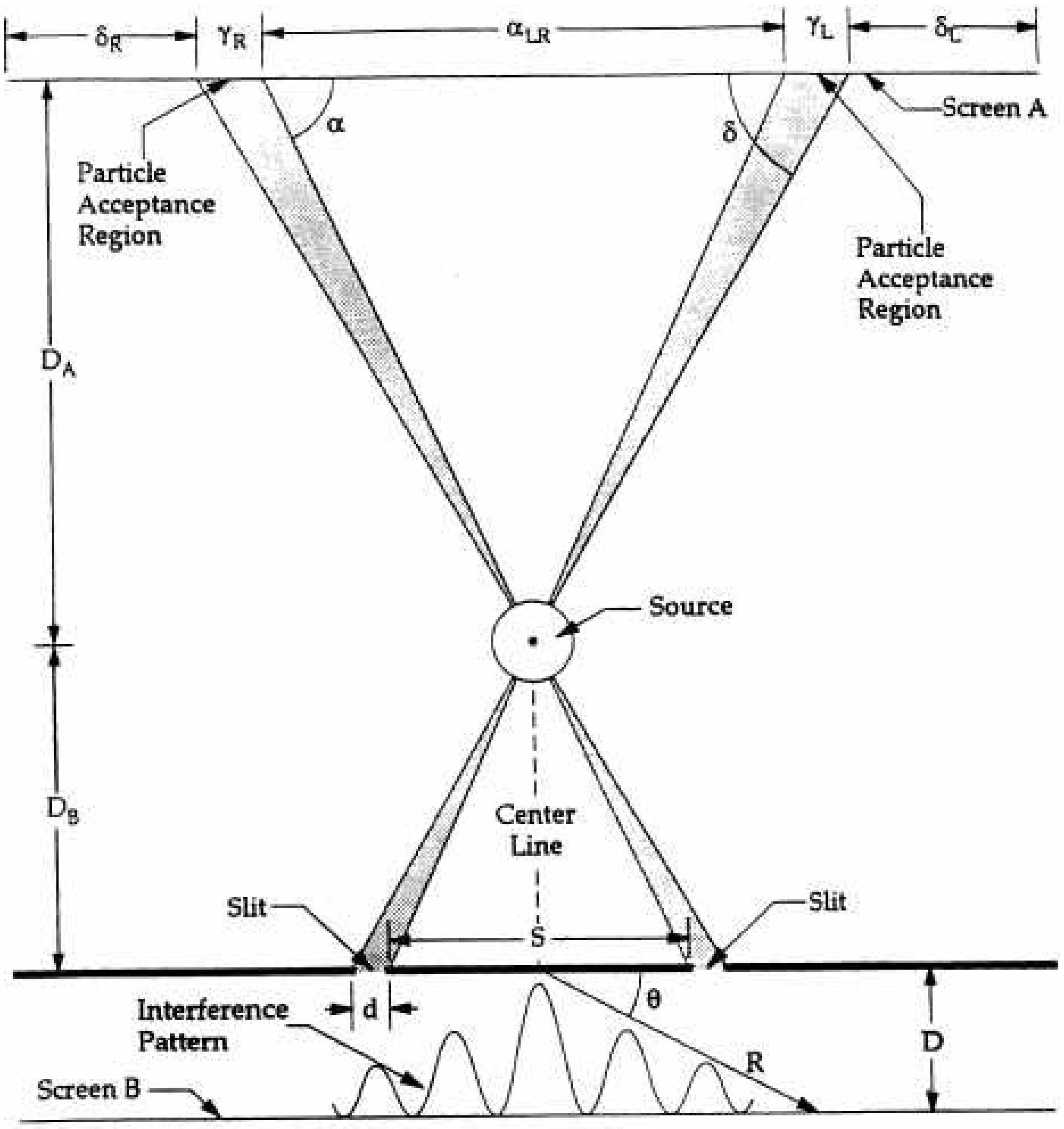

Fig. 1. Two-slit interference experiment (not drawn to scale) in which the traversed slit is determined by a partner (idler) particle emitted from the source in the opposite direction.

as determined by the projection of $\Delta x$ and the momentum uncertainty, in bringing particle B to the slit plate. Where s is the distance between the nearest edges of the slits.

Substituting eqs. (1) - (3) in (4), we obtain the quadratic equation

$$(\Delta x)^2 - s\Delta x + \frac{l D_B}{2p} < 0\,, \tag{5}$$

whose solution is

$$\Delta x < s\left[\tfrac{1}{2} \pm \tfrac{1}{2}\left\{1 - \frac{2 l D_B}{p s^2}\right\}^{1/2}\right]. \tag{6}$$

The uncertainty principle is violated if the second term of eq. (6) becomes imaginary. This is easily avoided, and our requirements are readily met if the wavelength

$$l < \frac{p s^2}{2 D_B}\ . \tag{7}$$

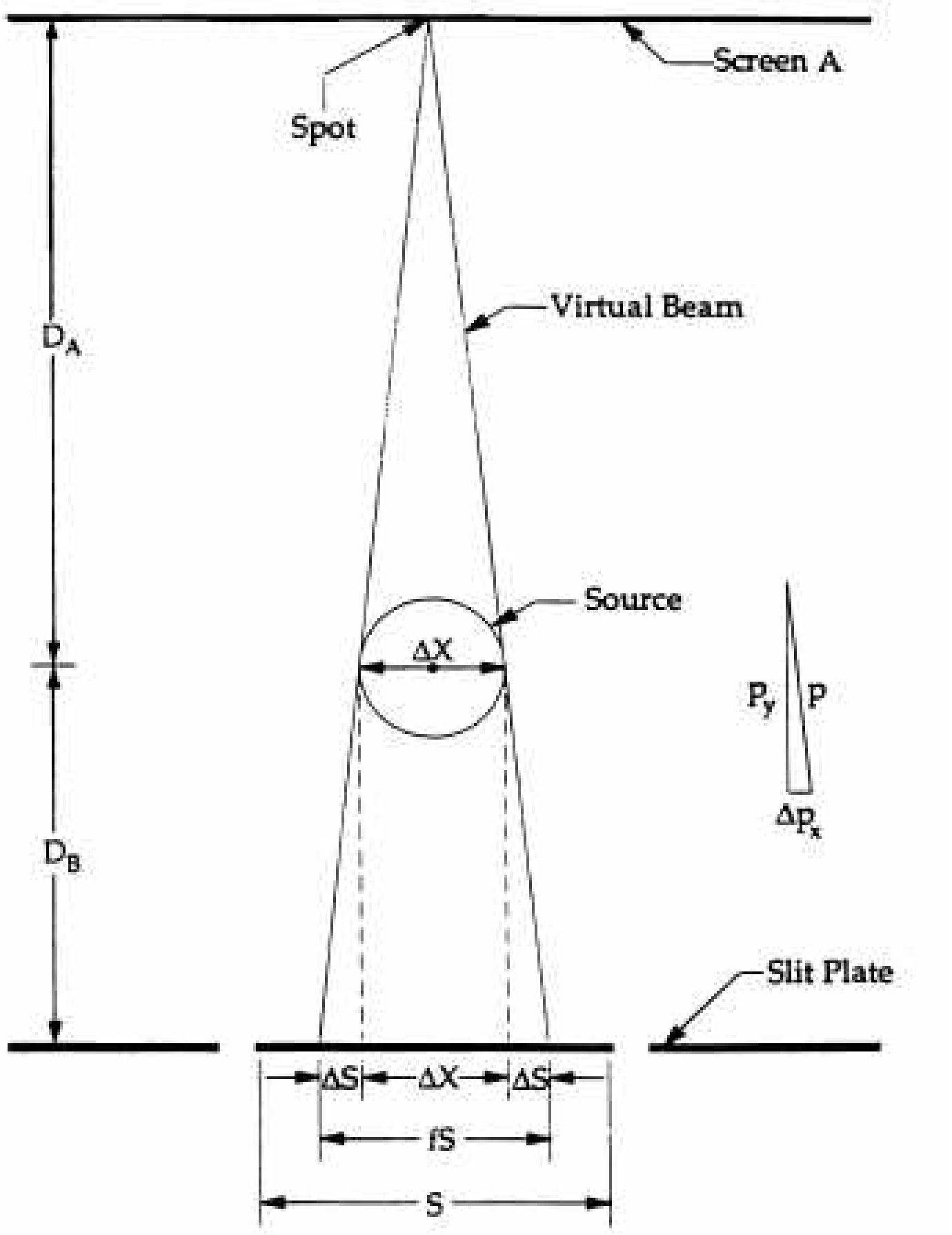

Fig. 2 . Example diagram (not drawn to scale) for analysis that the uncertainty principle is not violated in having a virtual beam of width fS (f is a number < 1), where S is the slit separation.

There is no great problem in meeting the condition given by eq. (7) to avoid a violation of the uncertainty principle. However, $l$ should not be made too small, as this would cause the interference fringes to be too close together to distinguish them. Violation of the uncertainty principle is avoided, since there is no need to measure momentum or velocity, but only position and direction.

Now that we have done an uncertainty principle analysis, it may be appropriate to ask if it would be possible to circumvent Heisenberg's uncertainty principle [16] for canonically conjugate variables by means of a comparable entangled particle stratagem. For example, could one in principle infer precise simultaneous momentum and position information about a particle moving in the + x direction. At a given instant in time (as measured in the lab frame) a momentum measurement $+p_{xo}$ can be made on a particle, while a position measurement of $-x_0$ is made on its entangled partner particle with which it is perfectly correlated. Can one then infer that at that instant a precise value $+x_0$ for the particle in addition to measuring its precise momentum $+p_{xo}$ ? In practice, the measurement errors would likely exceed the uncertainty principle errors.

## 5 Discussion of 1995 [9] "Which Slit" Gedanken Experiment

The quasi-stationary source emits a pair of particles in opposite directions by conservation of momentum, in which ideally the source had or gains little or no momentum perpendicular to the particles' trajectory. When this is not realized in practice, a partial correction can be computed taking into account the observed transverse component of momentum of the source. The source may emit a pair of photons resulting from a radiative cascade of Ca as in the experimental realization of the Einstein-Podolsky-Rosen-Bohm gedanken experiment by Aspect et al. , and in its modifications.[17 -19] Spontaneous parametric down conversion to prepare a source of entangled photons is another process. An additional process would be pair creation with equal and opposite momentum, such as electron-positron pairs.

Accumulative interference patterns for one-particle-at-a-time emission has been verified for particles such as photons, electrons, atoms, and neutrons [3, 20, 21]. In any case, two particles, A and B, are created in the source region simultaneously with equal and opposite momentum. Thus particles B are fired one at a time at the slit plate by repeating the emission process over and over. There is a coherence length constraint which limits the size of the source region so that the interference pattern is not lost due to phase differences between the particles as they are emitted sequentially.

Triangulation between the spot hit by particle A on screen A, the source, and a slit, classically determines whether particle B entered a given slit, or no slit at all. Figure 1 represents both point-like sources and extended sources where the particle-pair is emitted symmetrically with respect to the source center. As shown in Fig. 1, if the spot on screen A is in the particle acceptance region to the right of the axis of symmetry, particle B entered the left slit; and vice versa. Observation of the spot hit by particle A permits determination of the trajectory of particle B. If the spot on screen A is not in the particle acceptance region, then it could not have classically gone through either slit. Most classically allowed trajectories will miss both slits. Particle A carries mirror-image information of particle B's trajectory to the slit plate. For determination of emissions that miss the slits, the slit plate itself could be a luminescent screen.

## 6 Conclusion

### 6.1 Violation of Complementarity

By means of a position measurement on an entangled idler particle, it was possible to determine the traversed slit in a Young's two slit experiment, and still maintain the interference pattern. This violates Bohr's complementarity principle [4] for the wave-particle duality. It is likely that this will open the floodgates for violation of Bohr's complementarity principle in general. Varying parameters such as the distance from the source to the solitary screen, the distance from the source to the two-slit plate, the distance of the interference screen to the two-slit plate, the thickness of the two-slit plate, wavelength of the quantum particle, etc. may give further insight into the nature of entanglement.

### 6.2 de Broglie-Bohm Pilot-Wave

For the past six decades the Bohm modified pilot-wave theory of de Broglie, has been regarded as untestable with respect to the standard probabilistic interpretation of quantum mechanics. This is because both were regarded as always giving the same results, and Bohr's wave-particle complementarity principle was largely expected to be inviolable. Now that complementarity has been violated, not only is the pilot-wave theory separately testable, but a host of challenges may be raised with respect to the probabilistic interpretation of quantum mechanics. However, both theories are non-local. Although the pilot-wave theory answers many questions, it also raises questions of its own such as: what is the wave made of; does it interact with all particles in the same way; is it restricted to be a probability wave because of its non-locality; etc.? Even before the pilot-wave, Einstein once considered the electromagnetic field as the guiding wave for photons. Although violation of complementarity is a data point in favor of the pilot-wave theory, other theories may still be consistent with the known data.

### 6.3 Violation of Probabilistic Quantum Theory

Menzel et al [10] conclude that their findings do not contradict standard probabilistic quantum theory that asserts that quantum wave states are not objectively real until actually measured. But on p. 9314 they say: "This result is surprising because which-slit information about the signal photon is 'available' and therefore the principle of complementarity suggests no interference." Complementarity makes a much stronger statement in forbidding interference via an effectively collapsed signal photon vector state. If one of the wave functions collapses, this leaves no cross terms for interference. Contrary to their statement that this experiment demonstrates merely a surprising deviation of the probabilistic interpretation of quantum mechanics, it involves a major violation of that interpretation's basic principle. Saying that a body is in two mutually exclusive states at the same time until a measurement is made, is not the same thing as saying it is mathematically convenient to treat it as if the body were. Equal amplitude for both states can be assigned a priori or by a Bayesian approach.

A photon or particle in the de Broglie-Bohm theory must have real existence (must actually be present) in a particular state. This is why the probabilistic interpretation of a superposition state is not intrinsic to the pilot wave theory. One may consider a less stringent version of superposition as a statistical representation of finding a particle in one state – but not in both – whether or not a particular state is actually occupied.

The pilot-wave theory is inherently antithetical to complementarity as it is based on the separate coexisting reality of waves and particles. Furthermore, the pilot-wave theory can be viewed as unfettered by, if not antithetical to, the strict probabilistic interpretation of quantum mechanics. In the Menzel et al experiment [10], the waves on the two slits are certainly mutually coherent but there is not a superposition state of two separate photons present since we know through which slit the photon passed. Based on Menzel et al's experimental results, the respective "probabilities" on those two slits in a presumptive superposition state no longer represent the probabilities of finding the photon at both slits. The same may be true of the two $TEM_{01}$ lobes, but Menzel et al did not present sufficient data to decide this issue one way or the other as discussed in Sec. 3.3.